\documentstyle[aps,twocolumn,prl,epsf]{revtex}

\begin{document}

\newcommand{\be}{\begin{equation}}
\newcommand{\ee}{\end{equation}}
\newcommand{\bn}{\begin{eqnarray}}
\newcommand{\en}{\end{eqnarray}}

\draft

\twocolumn[\hsize\textwidth\columnwidth\hsize\csname @twocolumnfalse\endcsname

\title{Orbital Kondo Effect in $CrO_{2}$: A LSDA+DMFT Study}

\author{L. Craco, M. S. Laad and E. M\"uller-Hartmann}

\address{Institut f\"ur Theoretische Physik, Universit\"at zu K\"oln, 
Z\"ulpicher Strasse, 50937 K\"oln, Germany}
\date{\today}
\maketitle

\widetext

\begin{abstract}
Motivated by a collection of experimental results indicating the strongly 
correlated nature of the ferromagnetic metallic state of $CrO_{2}$, we 
present results based on a combination of the actual bandstructure~\cite{[3]} 
with dynamical mean-field theory (DMFT) for the multi-orbital case.  In 
striking contrast with LSDA(+U)~\cite{[3]} and model many-body 
approaches~\cite{[14]}, much better semiquantitative agreement with $(i)$ 
recent photoemission results, $(ii)$ domain of applicability of the half-metal 
concept, and $(iii)$ thermodynamic and $dc$ transport data, is obtained within 
a single picture.  Our approach has broad applications for the detailed first 
principles investigation of other transition metal oxide-based half-metallic 
ferromagnets.
\end{abstract}


]

\narrowtext

\begin{figure}[htb]
\epsfxsize=3.5in
\epsffile{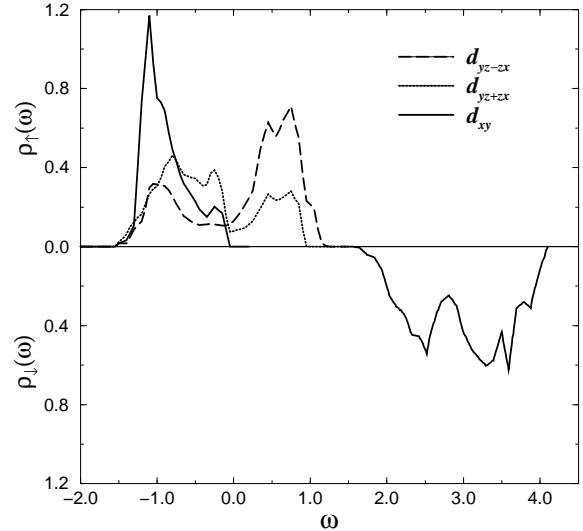}
\caption{$t_{2g}$ LSDA+U density of states per formula unit for both spin 
channels obtained from [3] , Figs. 3~(spin-down) and 4~(spin-up).}
\label{fig1}
\end{figure}

A lot of attention has recently been focused on $CrO_{2}$, widely used 
in magnetic recording, and with potential applications for 
spintronics~\cite{[1],[2]}. In contrast to the CMR manganites, stoichiometric 
$CrO_{2}$ is already a ferromagnetic metal.  Given the formal $4+$ valence 
state of $Cr$, the two $3d$ electrons occupy $t_{2g}$ orbitals. One would 
intuitively expect to form $S=1$ spin on each site, and an antiferromagnetic 
Mott insulator.  Why $CrO_{2}$ is a ferromagnetic metal instead, has been 
answered by Korotin {\it et al.}~\cite{[3]}, who have carried out insightful 
(LDA + U) calculations for this material.  Their main conclusions are: $(i)$ 
The triple-degeneracy of the $t_{2g}$ $d$-orbitals is lifted by tilting and 
rotation of the $Cr-O$ octahedra in the basic rutile structure, resulting in three bands with $xy, yz\pm zx$ 
character in the solid.  The $O$ $2p$ band(s) act, at least partially, as 
hole reservoirs resulting in $Cr$ being mixed-valent (like $Mn$ in doped 
manganites), explaining metallicity via self-doping in the negative 
charge-transfer situation realized in $CrO_{2}$. $(ii)$ an almost 
dispersionless majority spin band of 
predominantly $d$ character at about $1~eV$ below $E_{F}$ over a large region 
of the Brillouin zone.  This corresponds to strongly localized $xy$ orbitals 
completely occupied by one majority spin electron.  But the  
$d$ states (shown in Fig.~\ref{fig1}) of predominantly $yz\pm zx$ character 
hybridize with the $O$ $2p$ band and disperse, crossing $E_{F}$. The Hund's 
rule coupling between the localized $d_{xy}$ spin and the spin density of 
the band $d_{yz+zx}$ electrons polarizes the latter, giving a ferromagnetic 
state via the double-exchange (DE) mechanism.  Thus, both the metallicity and 
ferromagnetism are correlated well with each other.

Closer examination reveals that $CrO_{2}$ is a strongly correlated metal, 
implying that many-body correlation effects beyond the LDA~\cite{[4]} (or 
its variants) need to be considered. A number of experimental observations 
support such a view:

(1) Polarization dependent XAS measurements reveal substantial ligand orbital 
polarization.  An exchange splitting energy of $\Delta_{ex-spl}\simeq 3.2$ eV
was deduced~\cite{[5]}, implying substantial correlation effects, while 
LSDA calculations yield $\Delta_{ex-spl}\simeq 1.8 eV$!

(2) The resistivity has a characteristic correlated Fermi 
liquid (FL) form~\cite{[6]}: $\rho(T)=\rho_{0}+AT^{2}+BT^{7/2}$~\cite{[7]}, 
and, in fact, $CrO_{2}$ is a ``bad metal'' at high-$T$, with 
$\rho_{dc}(T>T_{c}^{FM}=390K)$ exceeding the Mott limit~\cite{[8]}. The 
Woods-Saxon ratio $A/\gamma^{2}$ is close to that expected for 
heavy-fermion metals~\cite{[8]}, implying substantial correlation effects 
in the Cr $d$-band.

(3) Optical conductivity studies reveal a Drude part at low energies, 
followed by a broad bump around $0.8~eV$ and high-energy features centered 
around $3~eV$. LDA+U predicts only the small Drude part 
correctly~\cite{[9]}. In addition, noticeable spectral weight transfer (SWT) 
from high to low energy is found as $T$ is reduced~\cite{[9]}. This SWT 
scales with magnetization, $M(T)$, as in the CMR 
materials~\cite{[10]}, showing clearly the correlated nature of the metallic 
state (the SWT is a {\it dynamical} many body correlation effect, and cannot 
be accessed by LSDA+U). 

(4) Recent measurements show that the integrated photoemission (PES) 
lineshape is characterized by a low-energy quasicoherent feature 
along with an incoherent~\cite{[11],[12]} broad feature at lower energies.  
This satellite feature observed in PES is a signal for the importance of 
dynamical, many-body correlation effects beyond LDA+U. 

(5) Finally, recent optical~\cite{[10]} and
 tunnelling measurements~\cite{[13]} show half-metallic character only
close to (about $\simeq 0.5~eV$ around) $E_{F}$.  Direct comparison with 
LSDA+U shows that these calculations would yield half-metallicity up to 
$1.5~eV$.It is not easy to cure this within LSDA+U.  One could, of course, 
refer to LSDA results, where the minority-spin band does have a threshold 
around $0.6~eV$, but LSDA gives results directly in conflict with (1)-(4).

The emergence of a correlated FL scale required to understand the above 
features is out of reach of LSDA  or pure DE models, because 
these are observed well below $T_{c}^{FM}$, and are thus related 
to additional scattering mechanisms in a half-metallic situation. We argued 
previously~\cite{[14]} that the above effects could be understood by invoking 
the important role of local, dynamical orbital correlations in the 
$t_{2g}$ sector. However~\cite{[14]}, a model (gaussian) density of states 
was used there, limiting direct comparison to experimental results. To do 
this, one has to extend a model-based approach to include real 
bandstructure features via the LDA+DMFT method. We choose LDA+DMFT because 
although approximations like LSDA+U do generate the correct ordered, 
{\it insulating} state(s), they fail to describe the correlated paramagnetic 
states, or the Mott transition accompanied by dynamical SWT. As discussed in 
detail~\cite{[15]}, very good agreement with experiment is achievable within 
LDA+DMFT.  In practice, this is a highly non-trivial task, and very 
computationally expensive. Here, we describe $CrO_{2}$ within this 
``first-principles bandstructure'' for correlated metals. The results are 
found to be in very good semiquantitative agreement with experiments probing 
both the occupied parts of the one-particle DOS ($\rho(\omega))$ as well as 
with thermodynamic and dc transport data cited above.   

An understanding of features mentioned above should go hand-in-hand with the
basic electronic structure.  Our starting point is the LSDA+U work of 
Korotin~{\it et al.}~\cite{[3]}, which yields the one-particle bandstructure 
of $CrO_{2}$ (notice that the Hartree-Fock shift is already incorporated in 
LSDA+U).  In Fig.~\ref{fig1}, we show the partial $t_{2g}$ DOS for the 
half-metallic situation realized in $CrO_{2}$.  In the real material, only 
the $t_{2g}$ states of $Cr$ hybridized with $O$-2p states are  
important, and so we neglect the higher energy $e_{g}$ bands.   
Following~\cite{[3]}, we suppose the $O$-2p states to introduce a small 
number of holes into the $d_{yz \pm zx}$ bands.  In fact, 
$\int_{-1.5}^{1.3}\rho_{\uparrow}^{t_{2g}}(\omega)d\omega=1.82<2$. 
Given that the $O$-2p spectral density is small compared to 
$\rho_{\uparrow}^{t_{2g}}(\omega)$, we keep only 
the $t_{2g}$ bands in what follows. With the LSDA bandstructure, the 
one-particle part of the Hamiltonian is:

\be
H_{0}=\sum_{{\bf k},\alpha,\beta}\epsilon_{\alpha\beta}({\bf k}) 
c_{{\bf k}\alpha\sigma}^{\dag}c_{{\bf k}\beta\sigma}
\ee
with $\rho_{\uparrow}^{t_{2g}}(\omega)=\sum_{{\bf k}\alpha\beta}
\delta(\omega-\epsilon_{\alpha\beta}({\bf k}))$ and 
$\alpha,\beta=xy,yz\pm zx$. To avoid double counting of Coulomb interaction 
contributions already contained in $H_{LDA}$, a term $H_{LDA}^{U}$ is 
subtracted from $H_{0}$.  Following~\cite{[15]}, the final result is,

\be
H_{0}=\sum_{{\bf k},\alpha,\beta}\epsilon_{\alpha\beta}({\bf k})
c_{{\bf k}\alpha\sigma}^{\dag}c_{{\bf k}\beta\sigma} 
+ \sum_{i\alpha\sigma}
\epsilon_{i\alpha\sigma}^{0} n_{i\alpha\sigma}\;,
\ee
where $\epsilon_{i\alpha\sigma}^{0}=\epsilon_{i\alpha\sigma}-
U(n_{\alpha\sigma}-1/2)+(1/2)J_{H}(n_{\alpha\sigma}-1)$, and $U$ and $J_{H}$
are defined below in $H$.

The full many body Hamiltonian for $CrO_{2}$ is:

\be
H=H_{0}+\frac{1}{2}\sum_{i\alpha\beta\sigma\sigma'}
U_{\alpha\beta}^{\sigma\sigma'}n_{i\alpha\sigma}n_{i\beta\sigma'}
-\frac{1}{2}\sum_{i\alpha\beta\sigma\sigma'}
J_{H}{\bf S}_{i\alpha}{\cdot}{\bf S}_{i\beta} \;.
\ee
 
To make progress, we notice that LSDA+U pushes the minority spin $t_{2g}$ 
bands to high energy $\omega>2$eV.  To begin with, we focus on the 
majority-spin $t_{2g}$ sector. A valid choice of parameters to describe 
this situation is given by the intra-orbital $U=5~eV$, and the Hund's rule 
coupling $J_{H}=1~eV$, close to the LSDA deduced values, along with the 
inter-orbital local interaction,$U'=U-2J_{H}$, for $t_{2g}$ orbitals. Notice 
that $U',J_{H}$ couple the three $xy, yz\pm zx$ bands, requiring 
an extension to treat multi-orbital effects.  

We solve this multiband hamiltonian in $d=\infty$ using the
iterated perturbation theory (IPT), suitably generalized for the multi-orbital
case.  Notice that $CrO_{2}$ falls into the class of TM oxides with 
well-separated $t_{2g}$ and $e_{g}$ bands~\cite{[3]}.  Moreover, 
 $G_{\alpha\beta\sigma\sigma'}(\omega)=
\delta_{\alpha\beta}\delta_{\sigma\sigma'}G_{\alpha\sigma} (\omega)$ and 
$\Sigma_{\alpha\beta\sigma\sigma'}(\omega)=
\delta_{\alpha\beta}\delta_{\sigma\sigma'}\Sigma_{\alpha\sigma} (\omega)$.

In the $t_{2g}$ sub-basis, a DMFT solution involves (i) replacing the 
lattice model by a self-consistently embedded multi-orbital, asymmetric 
Anderson impurity model, (ii) and a self-consistency condition which requires 
the local impurity Green function to be equal to the local GF for the lattice, 
given by 

\be
G_{\alpha}(\omega)=\frac{1}{V_{B}}\int d^{3}k \left[ 
\frac{1}{(\omega+\mu)1-H_{LSDA}^{0}({\bf k})-\Sigma(\omega)}
\right]_{\alpha} \;.
\ee

Using the locality of $\Sigma_{\alpha\beta}$ in $d=\infty$, we have
$G_{\alpha}(\omega)=G_{\alpha}^{0}(\omega-\Sigma_{\alpha}(\omega))$ using the 
Hilbert transform of the LSDA DOS.  In contrast to~\cite{[15]}, the 
inter-orbital coupling scatters electrons between the $xy,yz\pm zx$ bands,
so only the total number, 
$n_{d_{t_{2g}}}=\sum_{\alpha}n_{d_{t_{2g}},\alpha}$ is conserved in a manner 
consistent with Luttinger's theorem.  Further, in $CrO_{2}$, due to a highly 
asymmetric LDA DOS (Fig.~\ref{fig1}), we expect the final spectral function 
to reflect the interplay of the complicated bandstructure and inter-orbital 
correlations.

The calculation follows the philosophy of the IPT for the one-band Hubbard 
model, but the self-energies and propagators are matrices in the orbital 
indices~\cite{[16]}. Leaving details for a longer paper, we present here 
the final equations. The local propagators are given by,

\be
G_{\alpha}(\omega)=\frac{1}{N} \sum_{\bf k}
\frac{1}{\omega-\Sigma_{\alpha}(\omega)-\epsilon_{{\bf k}\alpha}} \;,
\ee
where $\alpha=xy, yz\pm zx$.  The local self-energies are computed from a 
generalized IPT formalism that takes into account the Luttinger theorem 
constraint off half-filling~\cite{[15]} (generalized Friedel sum 
rule). Explicitly, 

\be
\label{eq:self}
\Sigma_{\alpha}(\omega)=\frac{A_{\alpha}[\Sigma_{\alpha\alpha}^{(2)}(\omega)
+\Sigma_{\alpha\beta}^{(2)}(\omega)+\Sigma_{\alpha\gamma}^{(2)}(\omega)]}
{1-B_{\alpha}[\Sigma_{\alpha\alpha}^{(2)}(\omega)
+\Sigma_{\alpha\beta}^{(2)}(\omega)+\Sigma_{\alpha\gamma}^{(2)}(\omega)]}
\ee
where, for example, ($\alpha\ne\beta\ne\gamma$)

\be
\Sigma_{\alpha\beta}^{(2)}(i\omega)=
\left( \frac{U_{\alpha\beta}}{\beta} \right)^{2}
\sum_{n,m} G_{\alpha}^{0}(i\omega_{n})
G_{\beta}^{0}(i\omega_{m})G_{\beta}^{0}(i\omega_{n}+i\omega_{m}-i\omega)
\ee
and
$G_{\alpha}^{0}(\omega)=[\omega+\mu_{\alpha}-\Delta_{\alpha}(\omega)]^{-1}$
In Eqn.~(\ref{eq:self}), 
$A_{\alpha}=\frac{n_{\alpha}(1-2n_{\alpha})
+D_{\alpha\beta}[n]}{n_{\alpha}^0(1-n_{\alpha}^0)}$ and 
$B_{\alpha}=\frac{(1-2n_{\alpha})U_{\alpha\beta} +
\epsilon_{\alpha}-\mu_{\alpha}}
{2U_{\alpha\beta}^{2}n_{\alpha}^0(1-n_{\alpha}^0)}$.
Here, $n_{\alpha}$ and $n_{\alpha}^0$ are particle numbers determined from 
$G_{\alpha}$ and $G_{\alpha}^{0}$ respectively, and 
$D_{\alpha\beta}[n]=\langle n_{\alpha\sigma}n_{\beta\sigma'} \rangle$ 
is calculated using 
$\langle n_{\alpha\sigma}n_{\beta\sigma'} \rangle =
\langle n_{\alpha\sigma}\rangle \langle n_{\beta\sigma'} \rangle 
-\frac{1}{U'\pi}\int_{-\infty}^{+\infty}f(\omega)
[\Sigma_{\alpha}(\omega)G_{\alpha}(\omega)]d\omega$.
The last identity follows directly from the equations of motion for 
$G_{\alpha}(\omega)$.

\begin{figure}[htb]
\epsfxsize=3.5in
\epsffile{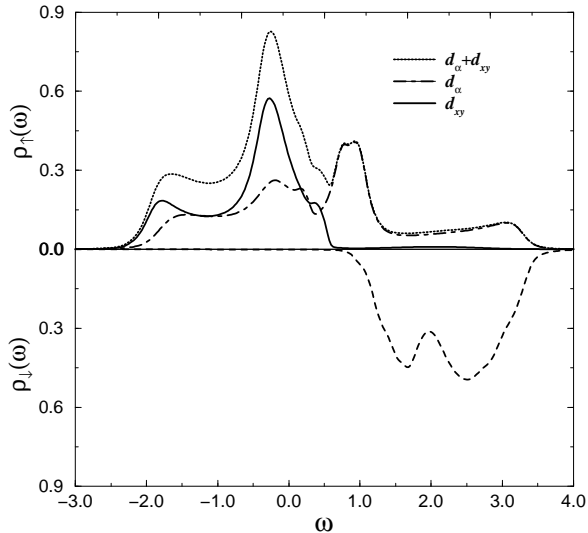}
\caption{LSDA+DMFT partial and total density of states for the $Cr$ $t_{2g}$ 
orbitals for $U'=3.0~eV$.}    
\label{fig2}
\end{figure}

We now present our results.  In Fig.~\ref{fig2}, we show the renormalized, 
total spectral function (sum of the partial $t_{2g}$ DOS) of $CrO_{2}$ for 
$T=0$. Notice the high-energy ``Hubbard bands'' appearing as manifestations
of dynamical effects of strong orbital correlations in the $t_{2g}$ sector.
As is also clear, the Luttinger theorem is obeyed to a very good accuracy.
Additionally, the self-energies show characteristic correlated Fermi liquid
behavior.  Based on these results, we interpret the low energy quasicoherent
features as arising from collective {\it orbital} Kondo screening of $t_{2g}$
orbital moments in the fully spin polarized half-metal.

We emphasize the quantitative differences between the results of 
this work with an earlier one~\cite{[14]}, where an idealized (gaussian) 
unperturbed DOS was used; no low-energy (LSDA related) pseudogap-like feature 
was visible there. Further, the $d_{xy}$ band was replaced there by a 
dispersionless, local level playing the role of polarizing the $yz\pm zx$ 
bands via strong $J_{H}$, making it impossible to access changes resulting 
from the interplay of the realistic bandstructure and strong orbital 
correlations.  These differences are especially important when one attempts to
describe spectroscopy results quantitatively (see below). 

\begin{figure}[htb]
\epsfxsize=3.5in
\epsffile{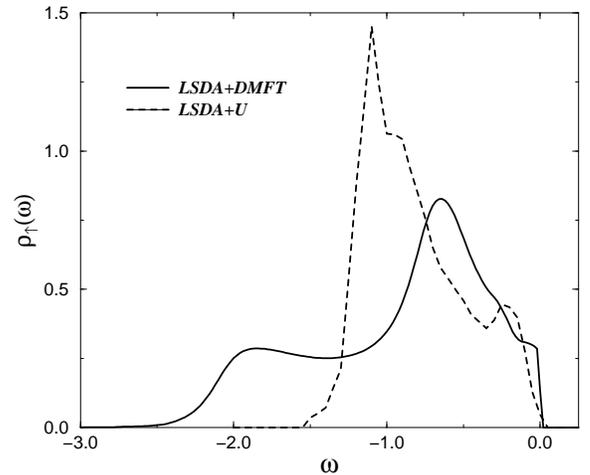}
\caption{The integrated photoemission lineshapes for the 
spin-up channel within LSDA+DMFT for $U'=3.0~eV$ (solid line) and 
LSDA+U (dashed line).} 
\label{fig3}
\end{figure}

In Fig.~\ref{fig3}, we show the integrated PES lineshape for low $T$, 
obtained directly from 
$I_{PES}(\omega)=f_{F}(\omega-E_F)\rho_{\uparrow}^{t_{2g}}(\omega)$.  Quite
satisfyingly, the quasicoherent spectral weight at $\mu$ is larger than what 
was reported in earlier experimental work~\cite{[11]}, and in quite good 
agreement with results obtained recently from thin films~\cite{[12]}.
In particular, it is gratifying to see good quantitative agreement between 
theory and experiment in the region $-1.2<\omega\le E_F$.  Specifically, the 
small, but clear dip seen in the LSDA+U DOS is absent in LSDA+DMFT, in 
agreement with the thin film results.  At higher binding
energy, we do not expect quantitative agreement because both $O$-2p and $e_{g}$
bands of $Cr$ begin to contribute, and also because the minority spin $t_{2g}$
band leads to a breakdown of our approach, which neglected the minority-spin
sector.  Consequently, further work is needed to make comparison with the 
recent tunnelling and optical measurements~\cite{[9],[13]}, since 
these measurements probe the minority spin band at higher energies.  Optical 
measurements reveal that the almost complete spin polarization near $E_{F}$ 
is reduced at higher energy $\omega \ge 0.5~eV$.  

To this end, we extend the above calculation to include the minority-spin
$t_{2g}$ bands, which (see Fig.~\ref{fig1}) are pushed to high energy 
$\omega > 2~eV$ by LSDA+U.  Inter-orbital correlations between the 
$t_{2g}\uparrow,\downarrow$ bands are accounted for in a way 
similar to that described above for the majority-spin sector.  However, a 
larger $U_{\uparrow\downarrow}=U'+2J_{H}$ is now required for the calculation 
of the minority spin DOS, $\rho^{t_{2g}}_{\downarrow}(\omega)$, alongwith
a Hartree shift of -$J_{H}m$ (where $m$ is the magnetization per site) 
for the $\downarrow$-spin sector. The interaction-corrected minority spin 
DOS (lower panel of Fig.~\ref{fig2}) shows interesting 
features.  First, $U_{\uparrow\downarrow}$ broadens out the sharp LSDA 
features, and more importantly, transfers dynamical spectral weight from 
high- to lower energy $\omega \simeq 0.7~eV$.  As a result, above $0.7~eV$, 
the spin polarization decreases continuously from saturation, in nice agreement
with indications from optical and tunnelling measurements~\cite{[9],[13]}. 
Notice that LSDA+U would predict complete half-metallicity up to a much higher 
energy $\simeq 1.5~eV$.  This represents strong evidence for the importance of 
treating dynamical SWT correctly, an effect missed by LSDA+U, but treated
adequately by LSDA+DMFT.  In fact, in our view, optical and tunnelling results 
provide a direct confirmation of the importance of dynamical correlation 
effects.  Indeed, one could argue that the near similarity of the LSDA- and 
LSDA+DMFT spectral functions for $t_{2g}^{\uparrow}$ sector for 
$-0.7\le\omega\le 0.2~eV$ is a sign of the adequacy of LSDA+U.  As discussed 
above, however, consistency with optics and tunnelling results as well is 
only achieved within LSDA+DMFT. 

 The results obtained above are also consistent with 
the low-$T$ thermodynamic and dc resistivity of $CrO_{2}$.  Indeed, from the 
computed self-energy, we estimate a moderate quasiparticle renormalization,
$Z$.  From the fact that Im$\Sigma_{yz\pm zx}(\omega)\simeq -b\omega^{2}$, 
we infer that the low-$T$ $dc$ resistivity should follow 
$\rho_{dc}(T)=\rho_{0}+AT^{2}$.  In fact, using 
$A=(m^{*}/ne^{2})(\partial^{2}\Sigma(\omega)/\partial\omega^{2})_{\omega=E_F}$
from $Im \Sigma_{yz\pm zx}(\omega)$, we estimate the Woods-Saxon ratio,
$A/\gamma^{2} \simeq 3.5~10^{-5}$, close to the value found 
experimentally~\cite{[8]}.

To conclude, we have extended the LSDA+U calculation~\cite{[3]} to explicitly
include the dynamical correlation effects arising from local electronic 
correlations in the $t_{2g}$ sector in the multi-orbital system $CrO_{2}$.
Puzzling signatures of strong correlations in the half-metallic state are 
understood as manifestations of collective orbital Kondo effect in the $t_{2g}$
sector, originating from a non-trivial interplay between the realistic hopping
in the rutile structure and inter-orbital ($t_{2g}$) correlations.
Much better quantitative agreement with PES is 
obtained with LSDA+DMFT than with LSDA or LSDA+U, showing clearly the
importance of dynamical correlation effects. Further, our results are 
also in semiquantitative agreement with thermodynamic and $dc$ transport 
measurements.  Essentially similar techniques can also be used fruitfully 
for other transition metal oxide-based ferromagnets currently of great 
interest~\cite{[17]}.

Work carried out under the auspices of the Sonderforschungsbereich 608 of the
Deutsche Forschungsgemeinschaft.

\end{document}